\documentclass[11pt,twoside]{article}

\usepackage{asp2006}
\usepackage{epsf}
\usepackage{psfig}
\usepackage{lscape}
\usepackage{amssymb,latexsym,graphicx,natbib,eufrak,times}

\def\gamma0{\hbox{$\Gamma_{\rm 0}$}}

\def\msun{\hbox{M$_\odot$}}
\def\mstar{\hbox{$M_\star$}}

\def\msunyr{\hbox{M$_\odot$\,yr$^{-1}$}}

\def\cm3{\hbox{cm$^{-3}$}}
\newcommand{\mmax}{M_{\rm max}}

\newcommand{\mup}{M_{\rm up}}
\newcommand{\lup}{L_{\rm up}}
\newcommand{\nup}{N_{\rm up}}

\newcommand{\mc}{M_*}
\newcommand{\lmax}{L_{\rm{max}}}

\newcommand{\dr}{\mbox{${\textup{d}}$}}

\newcommand{\dndmidt}{\frac{\dr N}{\dr M_i\dr t}}
\newcommand{\dndmidttext}{\dr N/(\dr M_i\dr t)}
\newcommand{\dndmdt}{\frac{\dr N}{\dr M\dr t}}
\newcommand{\dndmdttext}{\dr N/(\dr M\dr t)}

\newcommand{\cfr}{\textup{CFR}}
\newcommand{\cimf}{\textup{CIMF}}
\newcommand{\sfr}{\textup{SFR}}
\newcommand{\ml}{\Upsilon}
\newcommand{\dndldt}{\frac{\dr N}{\dr L\dr t}}
\newcommand{\dndldttext}{\dr N/(\dr L\dr t)}
\newcommand{\dndl}{\frac{\dr N}{\dr L}}
\newcommand{\dndltext}{\dr N/\dr L}

\newcommand{\dmidmtext}{\left|\partial M_i/\partial M\right|}

\begin{document}
\title{Basic Tools for Studies on the Formation and Disruption of Star Clusters: the Luminosity Function}   %%% Fill in title
\author{M. Gieles}   %%% Fill in author names
\affil{European Southern Observatory, Casilla 19001, Santiago, Chile}    %%% Fill in author affiliations

\begin{abstract} %%% Abstract to run on from here.
The luminosity function (LF) of young star clusters is often
approximated by a power law function. For clusters in a wide range of
galactic environments this has resulted in fit indices near $-2$, but
on average slightly steeper. A fundamental property of the $-2$ power
law function is that the luminosity of the brightest object ($\lmax$)
scales linearly with the total number of clusters, which is close to
what is observed. This suggests that the formation of Young Massive
Clusters (YMCs) is a result of the size of the sample, i.e. when the
SFR is high it is statistically more likely to form YMCs, but no
particular physical conditions are required. In this contribution we
provide evidence that the LF of young clusters is {\it not} a $-2$
power law, but instead is curved, showing a systematic decrease of the
(logarithmic) slope from roughly $-1.8$ at low luminosities to roughly
$-2.8$ at high luminosities.  The empirical LFs can be reproduced by
model LFs using an underlying cluster IMF with a Schechter type
truncation around $\mc\approx2\times10^5\,\msun$. This value of
$\mstar$ can not be universal since YMCs well in excess of this
$\mstar$ are known in merging galaxies and merger remnants. Therefore,
forming super massive clusters ($\gtrsim10^6\,\msun$) probably 
requires conditions different from those in (quiescent) spiral galaxies
and hence is not only the result of a size-of-sample effect.  From
the vertical offset a cluster formation efficiency of $\sim$10\% is
derived. We find indications for this efficiency to be higher when the
SFR is higher.
\end{abstract}

%%%%%%%%%%%%%%%%%%%%%%%%%%%%%%%%%%%%%%%%%%%%%%%%%%%%%%%
\section{Introduction: the luminosity function of young star clusters}
\label{sec:intro}

It has been well established that the luminosity function (LF) of
young star clusters is profoundly different from that of old globular
clusters (GCs). Where the latter is peaked (when presented in
logarithmic luminosity bins or in magnitude bins), the former rises
down to the detection limit.  When approximating the LF by a power law
function of the form

\begin{equation}
\dndl\propto L^{-\alpha}
\end{equation}
many studies find $\alpha\approx2$ for clusters in widely varying
environments \citep[for overviews see][]{2003MNRAS.343.1285D,
  2003dhst.symp..153W, 2006pces.conf...35L}.  However, the values
found for $\alpha$ are often slightly higher \citep[i.e the LF is
  steeper, see e.g.][]{2002AJ....123..207D, 2002AJ....123.1381E,
  2002AJ....124.1393L} and the range is rather large
($1.8\lesssim\alpha\lesssim2.8$) to be explained by statistical
fluctuations or photometric uncertainties.

The LF is often related to the cluster initial mass function
(CIMF). However, the relation between the LF and the $\cimf$ is not
one-to-one, since the LF consists of clusters of different ages and
evolutionary fading causes clusters of the same mass, but with
different ages, to contribute at different luminosities in the LF. The
empirically determined $\cimf$ seems also well described by a power
law, with an index of $-2$ \citep[e.g.][]{1997ApJ...480..235E,
  1999ApJ...527L..81Z, 2007ApJ...663..844M}, with the spread between
different studies much smaller than that found for the LF.

Although the LFs of the individual cluster systems are most of the
time within a few $\sigma$ compatible with a $-2$ power law, a
comparison between the different results shows that the deviations
from the $-2$ power law are not randomly scattered, but instead show
some systematic variations, in the sense that the LF is

\begin{enumerate}

\item {\it steeper at higher luminosities} \citep{1999AJ....118.1551W,
  2002AJ....123.1411B, 2002AJ....124.1393L, 2005A&A...443...41M,
  2006A&A...450..129G, 2006A&A...446L...9G, 2008AJ....135.1567H};

\item {\it steeper in redder filters}
  \citep[e.g.][]{2002AJ....123..207D,2002AJ....123.1381E,
    2006A&A...450..129G,2006A&A...446L...9G, 2008A&A...487..937H,
    2009arXiv0906.3432C}.
\end{enumerate}

\begin{figure}[!h]
\vspace{-0.75cm}
\center\includegraphics[width=8.5cm]{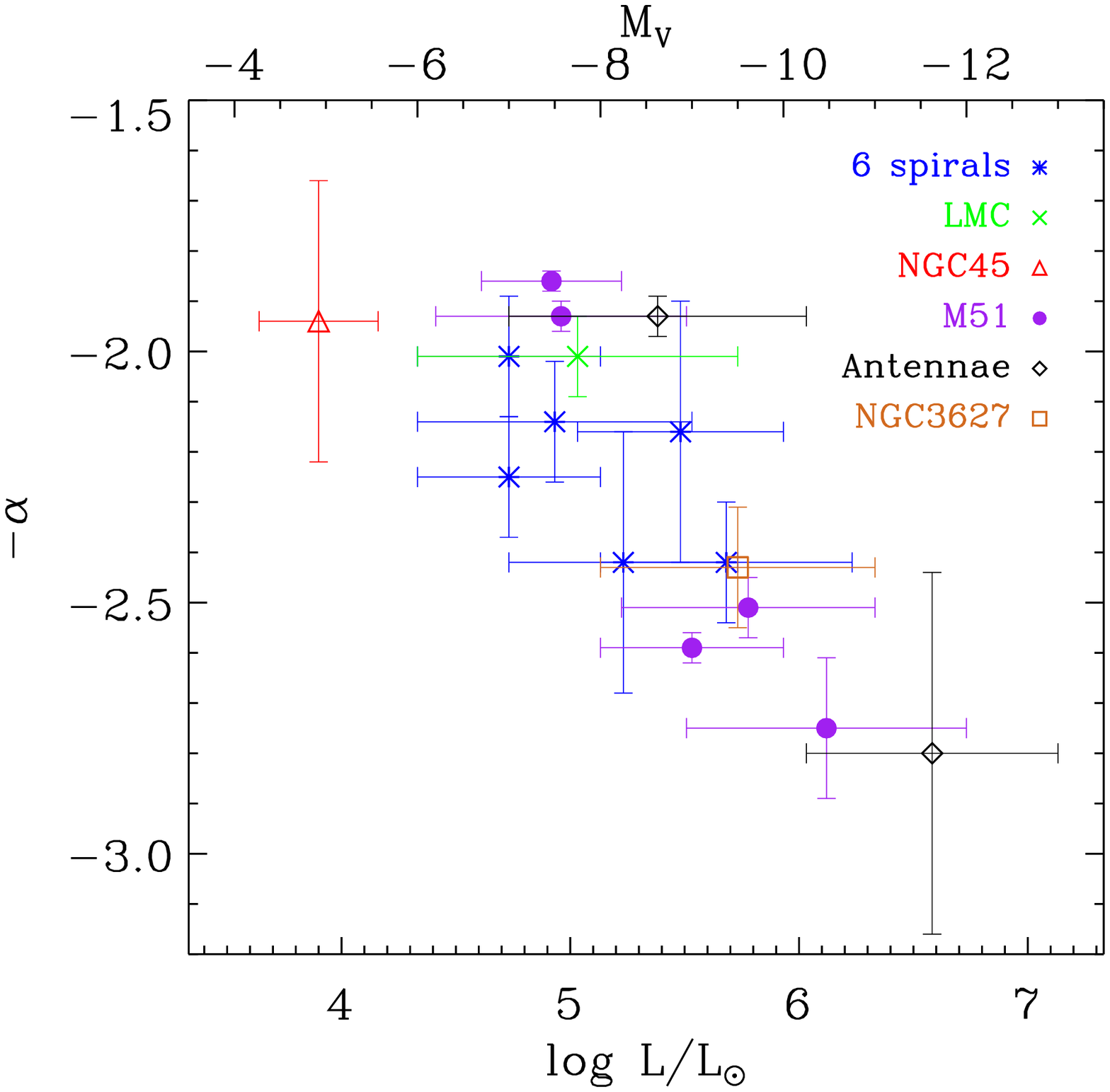}
\vspace{-0.5cm}
\caption{Sample of published indices of power law fit results to LFs of young
  star clusters as a function of the fit range. The results are taken
  from: \citet{2002AJ....124.1393L}: six spirals and the LMC;
  \citet{2007A&A...464..495M}: NGC~45; \citet{2006A&A...446L...9G,
    2008A&A...487..937H} and \citet{2008AJ....135.1567H}: M51;
  \citet{1999AJ....118.1551W}: Antennae; \citet{2002AJ....123..207D}:
  NGC~3627.}
\label{fig1}
\end{figure}

The steepening with increasing luminosity is illustrated in
Fig.~\ref{fig1} where we show results of power law fits to LFs of
clusters in different galaxies taken from literature. The horizontal
bars indicate the fit range used and the vertical bars give the
uncertainties in the power law index. From this a decrease of
$-\alpha$ (i.e. steepening) with increasing luminosity is seen. This
suggests that the $\cimf$ is not a universal continuous power law. If
it was a continuous power law with the same index at all ages and in
all galaxies, all LFs should be power laws with the same index. Age
dependent extinction or bursts in the formation rate would not cause a
difference between the CIMF and the LF. An addition of identical power
laws always results in the same power law. Luminosity dependent
extinction could cause a deviation from a $-2$ power law. This will be
addressed in Section~\ref{sec3}.

The fact that the LF gets steeper at the bright-end is an indication
that the CIMF is truncated at some upper mass $\mup$.  When an abrupt
truncation of the CIMF is assumed, it is possible to roughly estimate
the index of the bright-end of the LF. Assume that the CIMF is fully
populated, i.e. the mass of the most massive cluster actually formed,
$\mmax$, is equal to $\mup$.  Then assume a constant formation history
of clusters: $\dr\nup/\dr \tau=$constant. The luminosities of these
clusters, $\lup$, are age dependent. The light-to-mass ratio, or the
flux of a cluster of constant mass, scales roughly with age as
$\tau^{-\zeta_\lambda}$, with $0.6\lesssim\zeta_\lambda\lesssim1$
depending on the wavelength $\lambda$, such that
$\partial\lup/\partial\tau\propto\tau^{-\zeta_\lambda-1}\propto\lup^{1+1/\zeta_\lambda}$. The
same arguments hold for the luminosities of the
2$^\textup{\scriptsize{nd}}$, 3$^\textup{\scriptsize{d}}$, etc. most
massive clusters, such that the bright-end of the LF is

 \begin{eqnarray}
\frac{\dr N}{\dr L}({\textup{bright}})& \propto &\frac{\dr N}{\dr\tau}\left|
                                                  \frac{\partial\,\tau}{\partial L} \right|\\
	                              & \propto & L^{-1-1/\zeta_\lambda}.\label{eq:zeta}
 \label{eq:dndlindex}
\end{eqnarray}
For the $V$-band $\zeta_V=0.7$ \citep{2003MNRAS.338..717B,
  2007ApJ...668..268G}, such that for the $V$-band LF the index at the
bright-end is $-2.5$. For the $U(I)$-band $\zeta_{U(I)}\simeq1(0.6)$
and then equation~(\ref{eq:zeta}) predicts a logarithmic slope of
$-2(-2.7)$, i.e. shallower(steeper) at blue(red) wavelengths.  The
faint-end of the LF should still be a power law with index $-2$ at all
wavelengths or flatter if mass dependent disruption is important. This
double power law shape for the LF was found by
\citet{2006A&A...446L...9G} and \citet{1999AJ....118.1551W} for the
LFs of clusters in M51 and the Antennae, respectively.

In Section~\ref{sec2} we will model the LF in more detail and show
that also this double power law shape is not the best approximation of
the LF.

%%%%%%%%%%%%%%%%%%%%%%%%%%%%%%%%%%%%%%%%%%%%%%%%%%%%%
\section{Modelling the luminosity function}
\label{sec2}
In order to understand the luminosity function (LF) of star clusters
we create semi-analytic cluster population models. For this the
distribution function of initial masses ($M_i$) with age needs to be
defined first. We define the probability of forming a cluster with an
initial mass between $M_i$ and $M_i+\dr M_i$ at a time between $t$ and
$t+\dr t$, as $\dndmidttext$. For a \citet{1976ApJ...203..297S} type
CIMF this then gives
\begin{equation}
\dndmidt=A\,M_i^{-2} \exp(-M_i/\mc),
\label{eq:dndmisch}
\end{equation}
where $A$ is a constant that scales with the cluster formation rate
($\cfr$) and $\mc$ is the mass where the exponential drop occurs. Note
that we fix the power law index in the Schechter CIMF to $-2$.

An equivalent distribution function for the present day masses ($M$)
of clusters, $\dndmdttext$, can be found from multiplying
equation~(\ref{eq:dndmisch}) by $\dmidmtext$ which describes the
relation between the initial and the present cluster mass to taken
into account cluster dissolution. This was done by
\citet{2001ApJ...561..751F} and \citet{2007ApJS..171..101J} to model
the mass function of globular clusters using a constant mass loss
rate. \citet{2005A&A...441..117L} and \citet{2009MNRAS.394.2113G}
apply a mass dependent mass loss rate to model age and mass
distributions of cluster populations using a power law and a Schechter
CIMF, respectively. We refer the aforementioned studies and
\citet{2009A&A...494..539L} for full descriptions of the formulae.

An equivalent distribution function of luminosity ($L$) and $t$ can be
acquired by multiplying $\dndmdttext$ by the age dependent
mass-to-light ratio ($\ml$) \citep[see also][]{2006ApJ...652.1129F,
  2009A&A...494..539L}, such that

\begin{equation}
\dndldt = \dndmdt\ml(t).
\end{equation}
Since $\ml(t)$ needs to be taken from SSP models (we adopt the
\citet{2003MNRAS.344.1000B} models for solar metallicity and a
Salpeter stellar IMF) from here on the results need to be solved
numerically.

The luminosity function (LF) is then found by integrating
$\dndldttext$ over all ages

\begin{equation}
\dndl = \int \dndldt\dr t.
\end{equation}
The logarithmic slope at each $L$, which is a probe of the shape of
the LF, is found from
\begin{equation}
-\alpha(L) = \frac{\dr\log(\dndltext)}{\dr\log L}.
\label{eq:alpha}
\end{equation}

\begin{figure}
\vspace{-0.75cm}
\center\includegraphics[width=8.5cm]{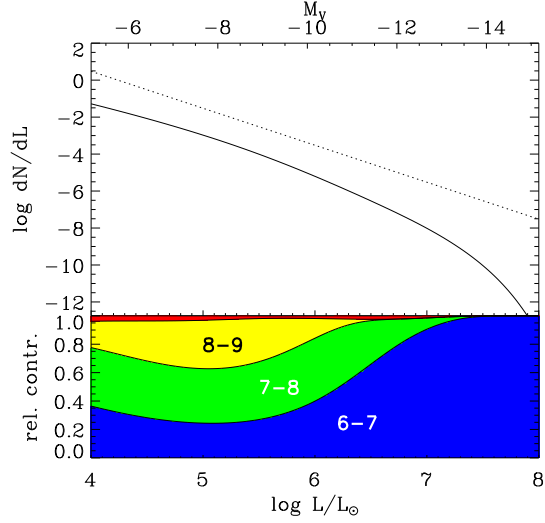}
\vspace{-0.5cm}
\caption{{\it Top panel:} Model for the LF based on a constant cluster
  formation rate, a Schechter cluster initial mass function with
  $\mc=2\times10^5\,\msun$ and a moderate cluster disruption time
  (full line). The dotted line shows a power law with index $-2$ for
  comparison. {\it Bottom panel:} the relative contribution of
  different age groups. The logarithmic age ranges are indicates. }
\label{fig2}
\end{figure}

An example of a model LF based on a Schechter CIMF and mass dependent
dissolution is shown in the top panel of Fig.~\ref{fig2}. The bottom
panel shows how clusters of different ages contribute to the LF. A
feature of the truncated CIMF is that at the bright-end mainly young
clusters contribute to the LF. This feature was indeed found from age
determinations of the brightest clusters in a sample of spiral
galaxies \citep{2009A&A...494..539L}.

%%%%%%%%%%%%%%%%%%%%%%%%%%%%%%%%%%%%%%%%%%%%%%%%%%%%%
\section{Comparison to the cluster populations of M51, M74 and M101}
\label{sec3}
Here the model of the previous section is compared to the LFs of star
clusters in 3 spiral galaxies: M51~(NGC~5194), M74~(NGC~628) and
M101~(NGC~5457). We use the data from \citet{2008A&A...487..937H,
  2008AJ....135.1567H}~(M51), \citet{2006AJ....132..883B}~(M101) and
Gieles~et~al.~(2010, in prep)~(M74).

First we take the model from Fig.~\ref{fig2} and apply a vertical
offset such that it matches the empirical LFs. This directly gives us
the cluster formation rate ($\cfr$) which enters in the model through
the variable $A$ in equation~(\ref{eq:dndmisch}). A comparison to the
global $\sfr$ in these galaxies (Table~1) shows that the cluster
formation efficiency
\citep[$\Gamma\equiv\cfr/\sfr$,][]{2008MNRAS.390..759B} is roughly
10\%, with a tendency of $\Gamma$ to increase with increasing
$\sfr$. Note that this fraction could be affected by disruption if
clusters are dissolving but are still picked up and therefore part of
the LF.

\begin{table}[!ht]
\caption{Comparison of the $\cfr$ derived from the LF of
  Fig.~\ref{fig2} to the $\sfr$ for the cluster populations considered
  here.}
\smallskip
\begin{center}
{\small
\begin{tabular}{ccccc}
\tableline
\noalign{\smallskip}
Galaxy & SFR & CFR & $\Gamma$\\
             & [$\msunyr$] & [$\msunyr$] &\\
\noalign{\smallskip}
\tableline
\noalign{\smallskip}
M74   &$\sim$1 &0.05 &0.05 \\
M101 &$\sim$3 & 0.25&0.08\\
M51   &$\sim$5 & 0.90& 0.18\\
\noalign{\smallskip}
\tableline
\end{tabular}
}
\end{center}
\end{table}

\begin{figure}[!t]
\vspace{-0.75cm}
\center\hspace{0.1cm}\includegraphics[width=7cm]{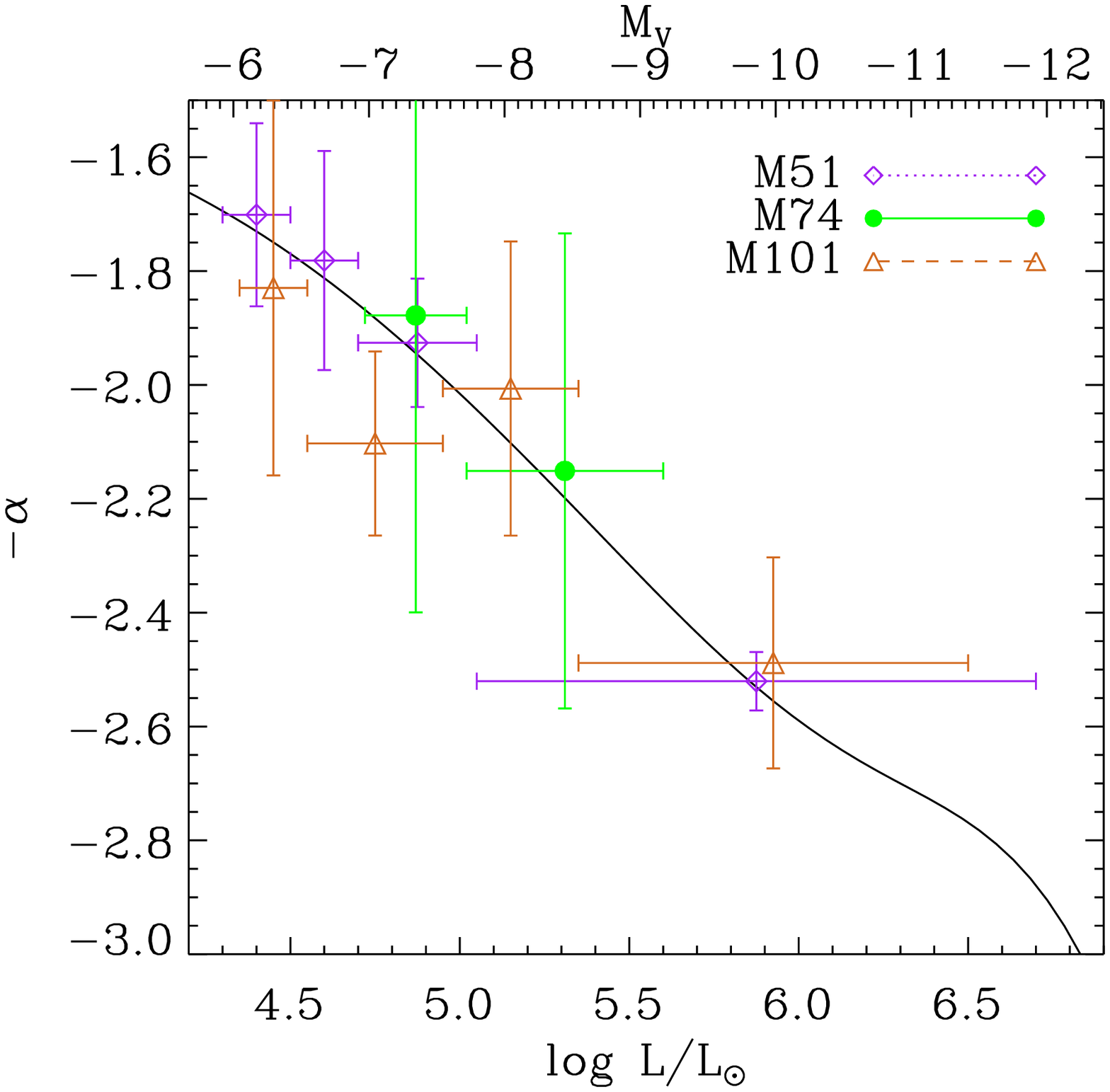}
\hspace{-1.5cm}\includegraphics[width=7cm]{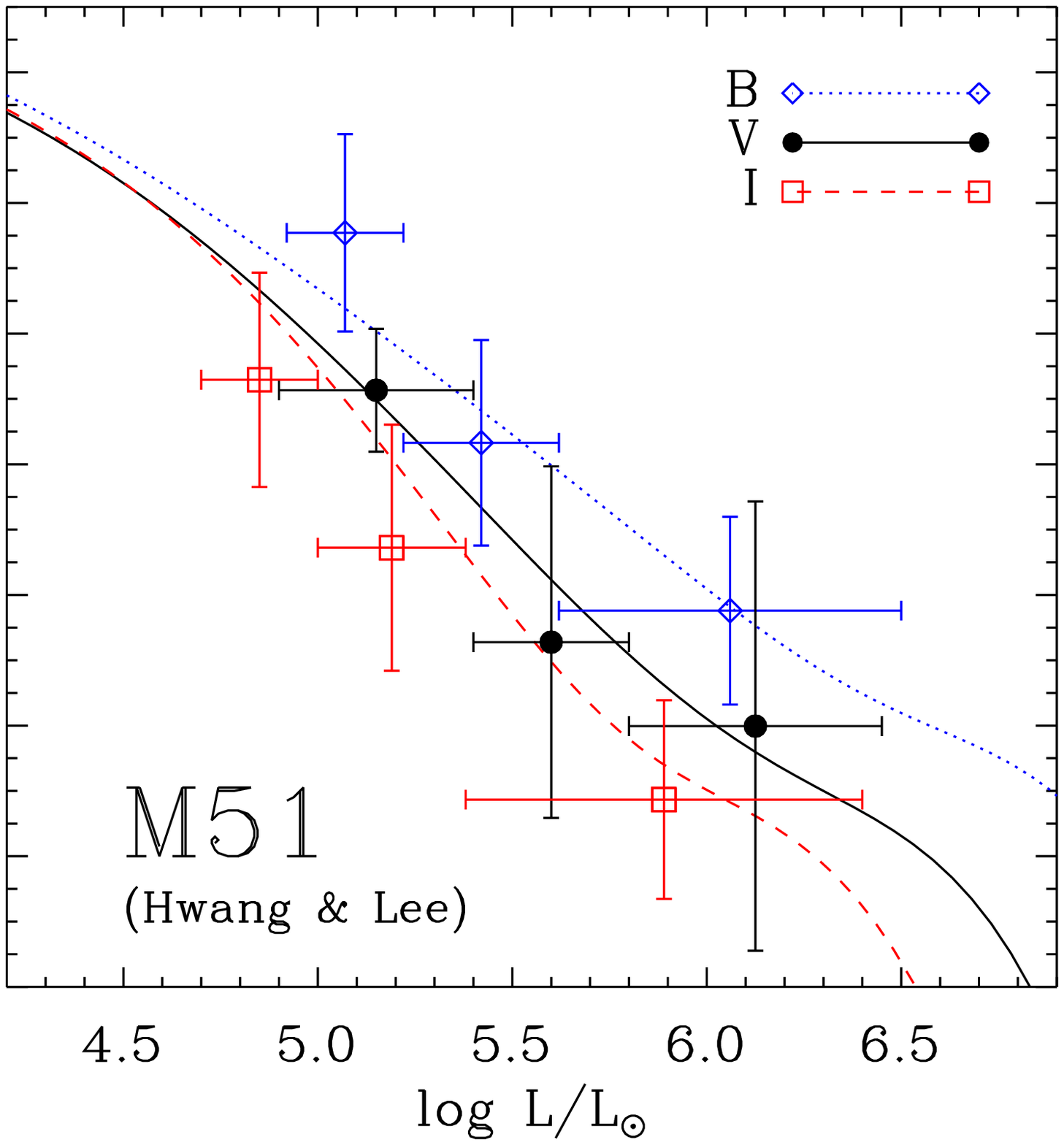}
\vspace{-0.5cm}
\caption{Logarithmic slope ($-\alpha$) vs. luminosity for the $V$-band
  LFs of clusters in M51, M74 and M101 ({\it left}) and for 3
  different filters for the M51 LF ({\it right}). The logarithmic
  slopes from the model shown in Fig.~\ref{fig2} (using
  equation~\ref{eq:alpha}) are overplotted.}
\label{fig3}
\end{figure}

Then we break the LFs in different luminosity segments and determine
the logarithmic slopes in these intervals using a maximum likelihood
estimate. The results are shown in Fig.~\ref{fig3}. As can be seen the
steepening of the LF with increasing $L$ as shown in Fig.~\ref{fig1}
is also found within the LF of individual cluster populations (left
panel). In the right panel we show that the filter dependent
logarithmic slopes for the M51 LF are also in good agreement with the
model.

So far we have ignored the effect of extinction on the individual
clusters. If luminous objects are more extincted, then the LF should
be steeper in the bluer filter. Since the opposite is found, it seems
that extinction does not play a vital role in shaping the LF.

%%%%%%%%%%%%%%%%%%%%%%%%%%%%%%%%%%%%%%%%%%%%%%%%%%%%%
\section{Summary}
\label{sec4}
We show that the LF of young star clusters is {\it not} a universal
$-2$ power law, but instead shows evidence for curvature, in the sense
that the logarithmic slope ($-\alpha$) decreases towards brighter
luminosities. This observational trend can be reproduced by a model
based on a Schechter type cluster initial mass function with
$\mstar\approx2\times10^5\,\msun$. A property of this model is that
$-\alpha(L)$ is smaller (steeper) in redder filters at the same $L$,
which is what we find for the rich cluster population of the
interacting galaxy M51.

%%% MAIN BODY OF TEXT GOES HERE. CONSULT "INSTRUCTIONS FOR AUTHORS USING
%%% LATEX2E MARKUP", SECTIONS 2.3-2.6 FOR HELP WITH EQUATIONS, FIGURES,
%%% AND TABLES.

%\section{}   %%% Top level section head (remove "%" symbol)
%\subsection{}   %%% Second level section head (remove "%" symbol)
%\subsubsection{}   %%% Lowest level section head (remove "%" symbol)
%\section*{}    %%% Unnumbered top level section head (remove "%" symbol)
%\subsection*{}   %%% Unnumbered second level section head (remove "%" symbol)

\acknowledgements 
MG thanks Pauline Barmby for kindly providing the luminosities of the
M101 clusters and the organisers of the meeting {\it Galaxy Wars:
  Stellar Populations and Star Formation in Interacting Galaxies} for
an enjoyable conference.

%%% THE BIBLIOGRAPHY
%%%
%%% CONSULT SECTION 3 OF "INSTRUCTIONS FOR AUTHORS" FOR HOW TO USE NATBIB.
%%% AUTHORS ARE ENCOURAGED TO USE EITHER THE "THEBIBLIOGRAPY" ENVIRONMENT
%%% BY UNCOMMENTING (DELETING THE "%" SYMBOL) THE COMMANDS BELOW, OR BY
%%% USING THE BIBTEX ENVIRONMENT. TO FIND OUT WHICH IS APPLICABLE TO YOUR
%%% CONTRIBUTION, CONSULT THE VOLUME EDITORS FOR YOUR PROCEEDINGS.
%%%

%\begin{thebibliography}{}
%\bibitem[]{}
%\bibitem[]{}
%\bibitem[]{}
%\bibitem[]{}
%\bibitem[]{}
%\bibitem[]{}
%\bibitem[]{}
%\bibitem[]{}
%\bibitem[]{}
%\bibitem[]{}
%\bibitem[]{}
%\bibitem[]{}
%\end{thebibliography}

\end{document}